\documentclass[%
 reprint,
 superscriptaddress,
nofootinbib,
 amsmath,amssymb,
 aps,
 prl
]{revtex4-2}

\usepackage{graphicx}%
\usepackage{dcolumn}%
\usepackage{bm}%

\usepackage{float}
\usepackage[dvipsnames]{color}
\usepackage[export]{adjustbox}
\usepackage{gensymb}
\usepackage{booktabs}
\usepackage[normalem]{ulem}
\usepackage{latexsym}
\usepackage{afterpage}

\makeatletter
\def\@fnsymbol#1{\ensuremath{\ifcase#1\or *\or \dagger\or \ddagger\or
   \mathsection\or \mathparagraph\or \|\or **\or \dagger\dagger
   \or \ddagger\ddagger \else\@ctrerr\fi}}
\newcommand{\ssymbol}[1]{^{\@fnsymbol{#1}}}
\makeatother

\usepackage{soul}

\newcounter{appendixctr}
\renewcommand{\theappendixctr}{\Alph{appendixctr}}

\usepackage{multirow}
\graphicspath{{figures/}} %
\usepackage{tikz}
\usepackage[normalem]{ulem}

\newcommand{\cmmnt}[1]{\ignorespaces}

\newcommand{\ket}[1]{\left|  #1 \right\rangle}

\newcommand{\blue}[1]{{\color{blue} {#1}}}

\usepackage[percent]{overpic}

\usepackage{hyperref}%

\begin{document}

\title{%
Loading and Imaging Atom Arrays via Electromagnetically Induced Transparency }

\author{Emily H. Qiu}
\thanks{These authors contributed equally}
 \affiliation{Department of Physics and Research Laboratory of Electronics, Massachusetts Institute of Technology, Cambridge, Massachusetts 02139, USA}

\author{Tamara \v{S}umarac}
\thanks{These authors contributed equally}
\affiliation{Department of Physics, Harvard University, Cambridge, Massachusetts 02138, USA}
\affiliation{Department of Physics and Research Laboratory of Electronics, Massachusetts Institute of Technology, Cambridge, Massachusetts 02139, USA}

\author{Peiran Niu}
\thanks{These authors contributed equally}
 \affiliation{Department of Physics and Research Laboratory of Electronics, Massachusetts Institute of Technology, Cambridge, Massachusetts 02139, USA}

\author{Shai Tsesses}
\affiliation{Department of Physics and Research Laboratory of Electronics, Massachusetts Institute of Technology, Cambridge, Massachusetts 02139, USA}

\author{Fadi Wassaf}
\affiliation{Department of Physics and Research Laboratory of Electronics, Massachusetts Institute of Technology, Cambridge, Massachusetts 02139, USA}
\affiliation{Department of Electrical Engineering and Computer Science, Massachusetts Institute of Technology, Cambridge, Massachusetts 02139, USA}

\author{David C. Spierings}
\affiliation{Department of Physics and Research Laboratory of Electronics, Massachusetts Institute of Technology, Cambridge, Massachusetts 02139, USA}

\author{Meng-Wei Chen}
\affiliation{Department of Physics and Research Laboratory of Electronics, Massachusetts Institute of Technology, Cambridge, Massachusetts 02139, USA}

\author{Mehmet T. Uysal}
\affiliation{Department of Physics and Research Laboratory of Electronics, Massachusetts Institute of Technology, Cambridge, Massachusetts 02139, USA}

\author{Audrey Bartlett}
\affiliation{Department of Physics and Research Laboratory of Electronics, Massachusetts Institute of Technology, Cambridge, Massachusetts 02139, USA}
\affiliation{Department of Chemistry, Massachusetts Institute of Technology, Cambridge, Massachusetts 02139, United States}

 \author{Adrian J. Menssen}
\affiliation{Department of Physics and Research Laboratory of Electronics, Massachusetts Institute of Technology, Cambridge, Massachusetts 02139, USA}

\author{Mikhail D. Lukin}
\affiliation{Department of Physics, Harvard University, Cambridge, Massachusetts 02138, USA}

\author{Vladan Vuleti\'{c}}
\affiliation{Department of Physics and Research Laboratory of Electronics, Massachusetts Institute of Technology, Cambridge, Massachusetts 02139, USA}

\begin{abstract}

Arrays of neutral atoms present a promising system for quantum computing, quantum sensors, and other applications, several of which would profit from the ability to load, cool, and image the atoms in a finite magnetic field. In this work, we develop a technique to image and prepare $^{87}$Rb atom arrays in a finite magnetic field by combining EIT cooling with fluorescence imaging. We achieve an average readout fidelity of $99.7(1)\,\%$ at $98.2(3)\,\%$ survival probability 
and up to $68(2)\%$ single-atom stochastic loading probability in a 2.3\,G magnetic field, with performance validated at fields up to 10\,G. We further develop a model to predict the survival probability, which also agrees well with several other atom array experiments. Our technique cools both the axial and radial directions, and will enable future continuously-operated neutral atom quantum processors and quantum sensors.

\end{abstract}

\maketitle

\section{I. INTRODUCTION}

Neutral atom arrays have emerged as a powerful platform for quantum technologies, enabling applications in quantum simulation~\cite{sepehr_quantumphases,scholl2021quantum,giulia_spinliquids}, quantum computation \cite{dolev_transport,Graham2022}, precision metrology~\cite{spinsqueezing_antoine, spinsqueezing_kaufman, spinsqueezing_monika}, and quantum-enhanced sensing~\cite{adams2019rydberg}. Such systems have demonstrated remarkable scalability compared to other platforms, routinely featuring hundreds to thousands~\cite{manetsch2025, pichard2024rearrangement} of individually trapped atomic qubits in optical tweezers. Even at these large scales, precise control is maintained over the atomic state and interatomic interactions, resulting in high-fidelity single-qubit rotations as well as two-qubit gate operations using excitation to Rydberg states~\cite{evered_high_fidelity,manuel_highfidelity,thompson_highfidelity,radnaev2024universal,senoo2025highfidelityentanglementcoherentmultiqubit}, which are on par with the state-of-the-art in other quantum processors. This high level of control, combined with reconfigurability in real time~\cite{dolev_transport}, has facilitated recent demonstrations of quantum information processing with logical qubits and error correction
techniques~\cite{dolev_logicalcomputer, bluvstein2025architecturalmechanismsuniversalfaulttolerant}, putting neutral atoms at the forefront of the quantum information landscape.

These experiments typically involve operation in a finite magnetic field, which enables long-lived storage and high-fidelity processing of quantum information~\cite{levine2019parallel}. However, these magnetic fields are not always compatible with preparation, cooling, and detection of individual atoms, which are needed for the operation of quantum sensors and processors based on atom arrays. Thus far, the magnetic field is switched on and off as required, however, the switching times for magnetic fields are significantly longer than other timescales of the quantum processor, limiting a variety of important operations ranging from mid-circuit readout~\cite{Graham2023midcircuit,Lis2023midcircuit,Norcia2023midcircuit} to continuous reloading~\cite{norcia2024iterative,zeiher_continuous,chiu2025continuousoperationcoherent3000qubit}. For operation of continuously-reloaded atom arrays, a constant, finite (non-zero) magnetic field is necessary for performing quantum operations on logical qubits, while ancilla atoms are simultaneously imaged and replenished from a continuously-reloaded atomic reservoir. Therefore, it is important to develop methods supporting the full functionality of atom arrays in magnetic fields compatible with quantum processing and sensing applications~\cite{sepehr_quantumphases,scholl2021quantum,giulia_spinliquids,dolev_transport,Graham2022,spinsqueezing_antoine, spinsqueezing_kaufman, spinsqueezing_monika,adams2019rydberg}.

For alkaline-earth and alkaline-earth-like species featuring narrow-linewidth transitions (such as Sr and Yb), Doppler cooling is possible in a finite magnetic field and is sufficient to reach temperatures necessary for high-fidelity qubit operations. Thus, loading and imaging of single atoms at finite magnetic fields have been demonstrated in such systems~\cite{norcia2024iterative, covey_yb_imaging,li2025fastcontinuouscoherentatom,Lis2023midcircuit}, paving the way towards their continuous operation~\cite{norcia2024iterative,zeiher_continuous,li2025fastcontinuouscoherentatom}. On the other hand, for alkali atoms (such as Rb and Cs), polarization gradient cooling (PGC) in zero magnetic field~\cite{dalibard1989} has been necessary to reach sufficiently low temperatures, thereby largely limiting preparation and readout processes to zero-field conditions. Although several alternative approaches to PGC have been demonstrated in the past~\cite{cindy_coolingimaging, lattice_D2Imaging,Li_GrayMolassesImaging,gadway2022,Haller2015}, as well as nondestructive imaging with PGC in finite magnetic fields~\cite{pritchard_prl_2023, saffman2017_state_selective}, these were performed either with significantly deeper traps, with significantly longer timescales, or with insufficiently high magnetic fields for qubit operation. The different experimentally-demonstrated cooling methods for various atomic species and the associated parameters, including photon scattering rates and magnetic fields, are summarized in Table~\ref{tab:cooling} of Appendix~\ref{app:imaging_review}.

In this work, we demonstrate a method for single-atom loading and high-fidelity, nondestructive imaging of a $^{87}$Rb atom array, in finite magnetic fields of up to $10$\,G.
Our scheme combines three-dimensional laser cooling based on Electromagnetically Induced Transparency (EIT)~\cite{morigi2000,kurtsiefer2024} with separately adjustable, simultaneous fluorescence imaging. The photon scattering rate achieved using this method ($\Gamma_{sc}\approx 9 \times 10^4\,$photons/s, limited by the trapping frequency)
is comparable to that used for zero-field imaging (see Appendix~\ref{app:imaging_review} and Table~\ref{tab:cooling}). Despite the modest imaging-related parameters of our apparatus\footnote{Microscope objective lens (NA = 0.4 and $60\%$ imaging transmission), tweezer parameters ($w_0 = 1.6\,\mu$m, $\omega_r/(2\pi)= 45$\,kHz and $\omega_z/(2\pi)= 4$\,kHz) and atom lifetime ($2.3$\,s)}, we achieve $99.7(1)\,\%$ readout fidelity and $98.2(3)\,\%$ atom survival probability
in $70$\,ms (brackets indicate standard error of the mean across traps). With a state-of-the-art vacuum lifetime and microscope objective, an order-of-magnitude speedup in imaging, as well as higher survival probability, are expected (see Appendix~\ref{app:speedup} and Table~\ref{tab:speedup}). In addition, we show that our method can be used for single-atom loading within $10$\,ms, and that the loading efficiency can be enhanced to $68(2)\,\%$ by using blue-detuned light. Compared to very recently demonstrated one-dimensional PGC in a finite magnetic field~\cite{bluvstein2025architecturalmechanismsuniversalfaulttolerant, chiu2025continuousoperationcoherent3000qubit}, our beam geometry cools more efficiently in three dimensions, supporting fast atom reuse in quantum processors and sensors~\cite{bluvstein2025architecturalmechanismsuniversalfaulttolerant}.

In addition, we develop a model based on collisions between excited cold atoms and room-temperature background atoms to explain the observed atom loss during imaging, reaching order-of-magnitude agreement across various atom array experiments as well as for our own. Notably, our model implies that improvements in vacuum lifetime and/or removal of the atomic background vapor can reduce imaging-induced atom loss, which could enable even larger future neutral-atom quantum processors.

\begin{figure}[htbp]
\centering
\includegraphics[width=0.48\textwidth, trim= 180 134 230 85,clip]{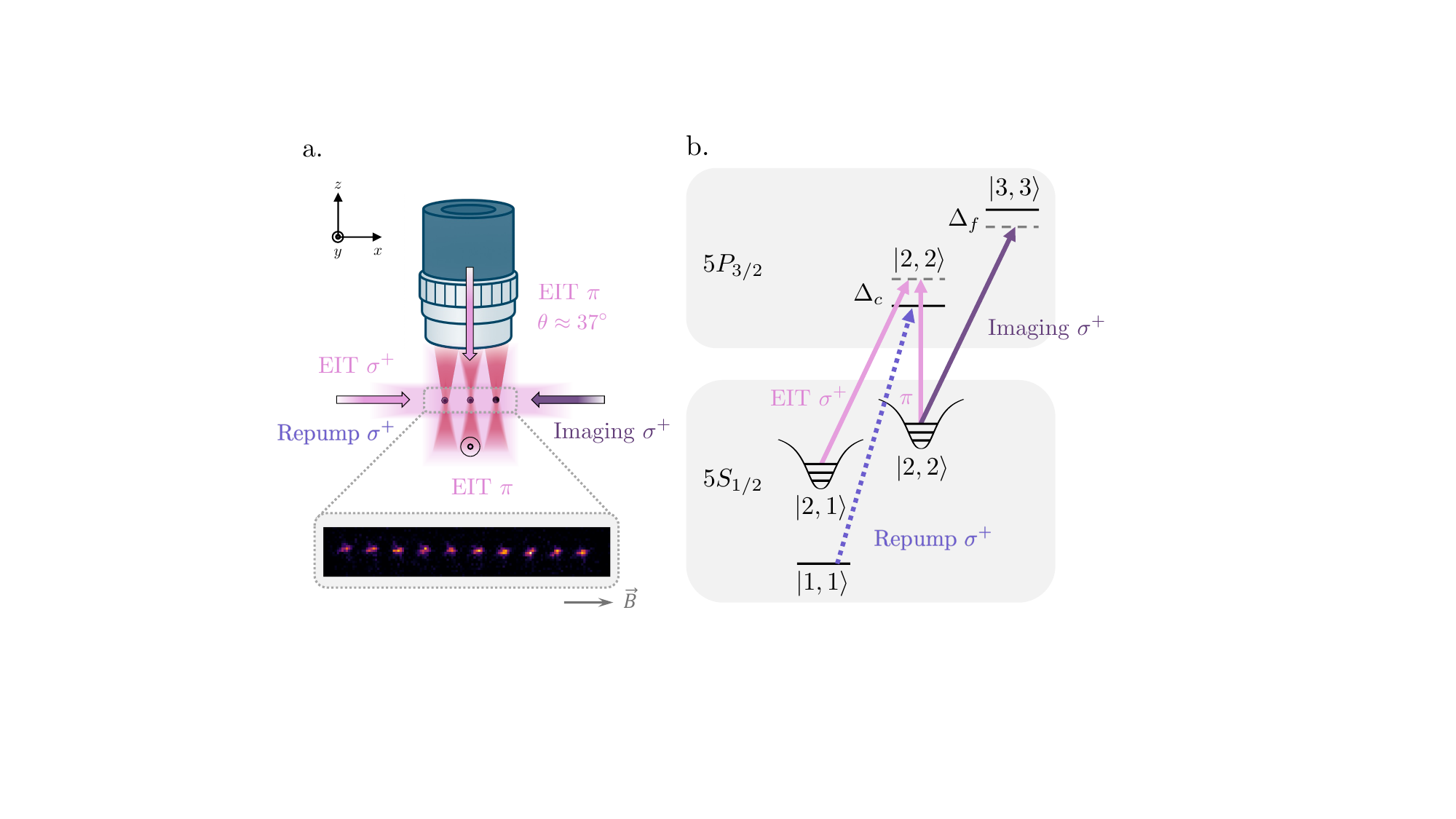}
\caption{\textbf{Experimental setup and scheme for loading and imaging tweezer arrays at finite magnetic fields: a.} Laser beam configuration (see text) and $10$--atom tweezer array imaged in a 2.3~G magnetic field oriented along the $x$--axis.
Our scheme cools both the axial and radial axes of the tweezer. 
\textbf{b.} Energy level diagram for EIT cooling and simultaneous imaging.
}
\label{fig:Setup}
\end{figure}

\begin{figure*}[htbp]
\centering
\includegraphics[width=\textwidth]{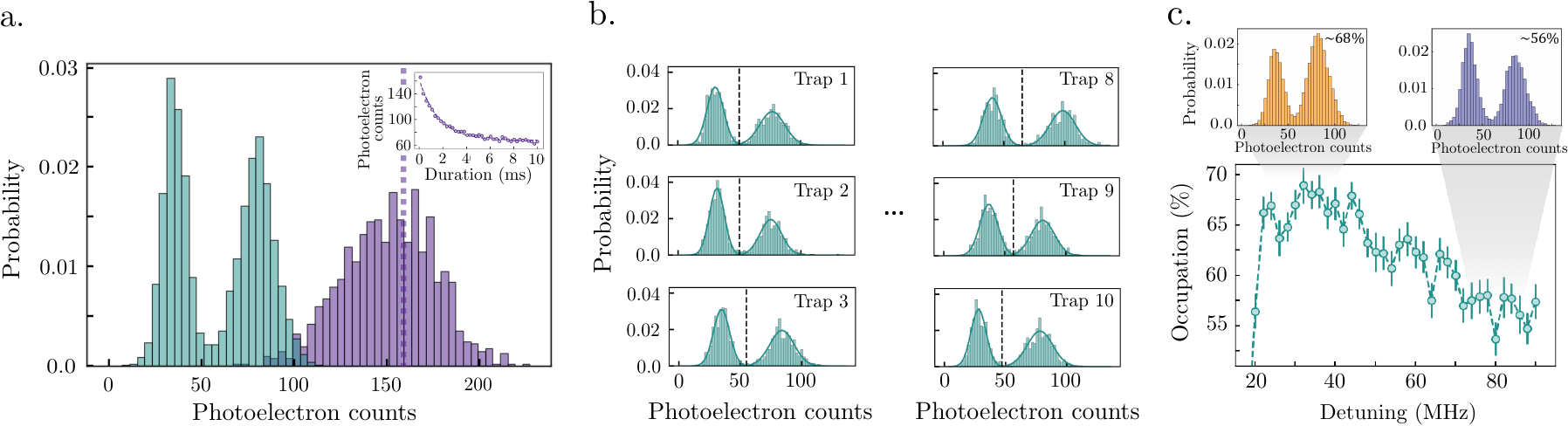}%
\caption{ \textbf{Loading and imaging of atoms in a finite magnetic field: a.} Histograms of two consecutive EIT images (each of $70$~ms duration) in a 2.3~G magnetic field. The first image (purple histogram) with a single, broad distribution shows that ensembles of atoms are initially loaded into the tweezer. The second image (green histogram) displays the characteristic bimodal distribution confirming that single atoms have been prepared stochastically through light-assisted collisions during the first image. \textit{Inset:} Photoelectron counts observed during the second image (fixed) as a function of the first image duration. The fit reveals a characteristic light-assisted collision time of $1.63(7)$~ms, with single atoms reliably prepared within $10\,$ms.  \textbf{b.} Histograms of EIT images in an array of 10 traps show an average detection fidelity of $99.7(1)\,\%$
with a stochastic loading probability of $49(1)\,\%$.
\textbf{c.} Single-atom preparation probability as a function of imaging beam detuning from the Stark-shifted $\ket{5S_{1/2}, F=2} \rightarrow \ket{5P_{3/2}, F'=3}$ transition. We observe $68(2)\%$ enhanced loading probability (orange histogram) for detunings $\lesssim 2U/h = 42\,$MHz, as expected~\cite{brown_gray_molasses}. Brackets indicate standard error of the mean across all traps.
}
\label{fig:Imaging and Loading}
\end{figure*}

\section{II. EXPERIMENTAL SETUP}

Our experimental setup is shown in Fig.~\ref{fig:Setup}a. A spatial light modulator (SLM) is used to generate an array of 10 optical tweezers (trap wavelength $\lambda = 808$\,nm) focused to beam waists of $w_0 \simeq 1.6\,\mu$m through a microscope objective lens ($\textrm{NA}=0.4$). The measured radial and axial trap frequencies are $\omega_r/(2\pi)= 45$\,kHz and $\omega_z/(2\pi)= 4$\,kHz, respectively, at a trap depth of $U/h=18.3\,$MHz.
After initial magneto-optical-trap (MOT) cooling, loading, and a MOT compression stage optimized with machine learning~\cite{Wigley2016mloop,our_bec_paper}, an ensemble of $^{87}$Rb atoms is prepared in the tweezer region. The magnetic gradient is then switched off, while the bias field along the $x$ direction is applied, typically to $2.3$~G (see Appendix~\ref{app:setup} for details). We emphasize that the current experiment was performed as a proof-of-principle for preparing and imaging single atoms in a finite magnetic field, and the operation of future, continuously-reloaded atom arrays will not require any magnetic field switching (see Appendix~\ref{app:application}).

The operations of single-atom loading and imaging are performed with a combination of EIT cooling beams and a fluorescence imaging beam. Without the fluorescence imaging beam, photon scattering from the EIT cooling beams quickly stops as atoms accumulate in the dark state~\cite{cindy_coolingimaging}. The EIT cooling beams consist of a $\sigma^+$--polarized control beam used for optical pumping (propagating along the magnetic field direction, i.e. the $x$--axis), and a $\pi$--polarized probe beam (aligned in the $y$--$z$ plane at an angle of $37^\circ$ relative to the $z$--axis) for driving two-photon Raman transitions that reduce the atoms' kinetic energy. The fluorescence imaging beam is $\sigma^+$--polarized, counterpropagating to the EIT control beam. An additional $\pi$--polarized EIT probe beam aligned along the $y$--axis was found to enable higher fluorescence scattering rates for fixed atom loss, but it is not strictly necessary for detection, making our approach compatible with a zoned architecture for quantum computing with neutral atoms~\cite{dolev_transport}. 

In Fig.~\ref{fig:Setup}b, the relevant Zeeman sublevels used for imaging and cooling are shown. The fluorescence imaging beam is detuned by $\Delta_f/(2\pi)=-\,67\,$MHz from the $\ket{5S_{1/2}, F=2} \rightarrow \ket{5P_{3/2}, F'=3}$ transition for detection. The $\sigma^+$ EIT control beam is $\Delta_c/(2\pi)=+\,36\,$MHz detuned from the $\ket{5S_{1/2}, F=2} \rightarrow \ket{5P_{3/2},F'=2}$ transition, and the $\pi$--polarized EIT probe light frequency is tuned to drive Raman transitions between the $m_F=2$ and $m_F=1$ magnetic sublevels in the $F=2$ ground state~\cite{kurtsiefer2024}. A circularly polarized repumping beam addresses the $\ket{5S_{1/2}, F=1} \rightarrow \ket{5P_{3/2}, F'=2}$ transition, generated as a sideband on the $\sigma^+$ control beam via a phase electro-optic modulator (EOM). For more details on experimental parameters including Rabi frequencies, and a calculation validating that we operate outside the Lamb-Dicke regime, see Appendix~\ref{app:setup}.

\section{III. RESULTS}

In Fig.~\ref{fig:Imaging and Loading}a, we show the preparation of single atoms in optical tweezers in a finite magnetic field, for a typical trap within the array. The histogram in purple displays the photon counts collected during $70$~ms of imaging in a 2.3~G field, performed immediately after tweezer loading from the MOT. This signal corresponds to the fluorescence from several trapped atoms as they undergo light-assisted collisions. The histogram in green shows the photon counts collected from a subsequent image, which displays a bimodal distribution and confirms stochastic loading of single atoms. The average loading probability is $49(1)\,\%$, 
consistent with typical loading probabilities during PGC in zero magnetic field~\cite{first_singleatom, andersen_review_2016}. The inset shows that the timescale to prepare single atoms is much shorter than the imaging duration required with our microscope objective. We measure a characteristic collision timescale of $1.63(7)$~ms, with single atoms prepared reliably within $10$~ms.

Fig.~\ref{fig:Imaging and Loading}b shows histograms of subsequent images with zero or one atom in an array of 10 traps, from which we extract an average fidelity of $99.7(1)\,\%$
to distinguish a single atom from the background (see Appendix~\ref{app:imaging_details} for details on performance for each trap). During $70$~ms of imaging, we detect on average 100 photons (50 photoelectrons) per atom, %
consistent with our estimated imaging beam scattering rate  combined with our estimated collection ($3.3\,\%$) and overall detection efficiencies ($\approx 1\,\%$), which is predominantly limited by our microscope objective. With a reasonable upgrade of our objective, we expect our imaging duration to be substantially reduced to $\lesssim 10$~ms (using an objective with NA = 0.65, $90\%$ transmission; see Appendix~\ref{app:speedup})~\cite{zhang_high_optical_regal, sepehr_thesis, chiu2025continuousoperationcoherent3000qubit}. 

For our EIT cooling scheme~\cite{morigi2000}, the cooling beams are blue-detuned relative to the $\ket{5S_{1/2}, F=2} \rightarrow \ket{5P_{3/2}, F'=2}$ transition, which simultaneously enables enhanced loading via blue-detuned light-assisted collisions~\cite{brown_gray_molasses, lester_rapid_2015, grunzweig2010}. In Fig.~\ref{fig:Imaging and Loading}c, we demonstrate enhanced single-atom loading probabilities of up to $68(2)\,\%$ by changing the detuning of the fluorescence imaging beam $\Delta_f$ from red- to blue-detuned relative to the $\ket{5S_{1/2}, F=2} \rightarrow \ket{5P_{3/2}, F'=3}$ transition. Enhanced loading is observed for detunings $\lesssim 2U/h = 42\,$MHz relative to the Stark-shifted line, as expected~\cite{brown_gray_molasses, lester_rapid_2015, grunzweig2010}. Moreover, our characteristic single-atom preparation timescale for enhanced loading is only $2-10\,$ms, an order of magnitude shorter than previous demonstrations that used the $D_1$ line~\cite{brown_gray_molasses, lester_rapid_2015, grunzweig2010}. While $\Lambda$-enhanced gray molasses cooling has been previously demonstrated on the $D_2$ line~\cite{Rosi2018}, we find that it is also possible to simultaneously enhance the probability of preparing single atoms, which eliminates the need for additional $D_1$ lasers.

\begin{figure*}[htbp!]
\centering
\includegraphics[width=\textwidth]{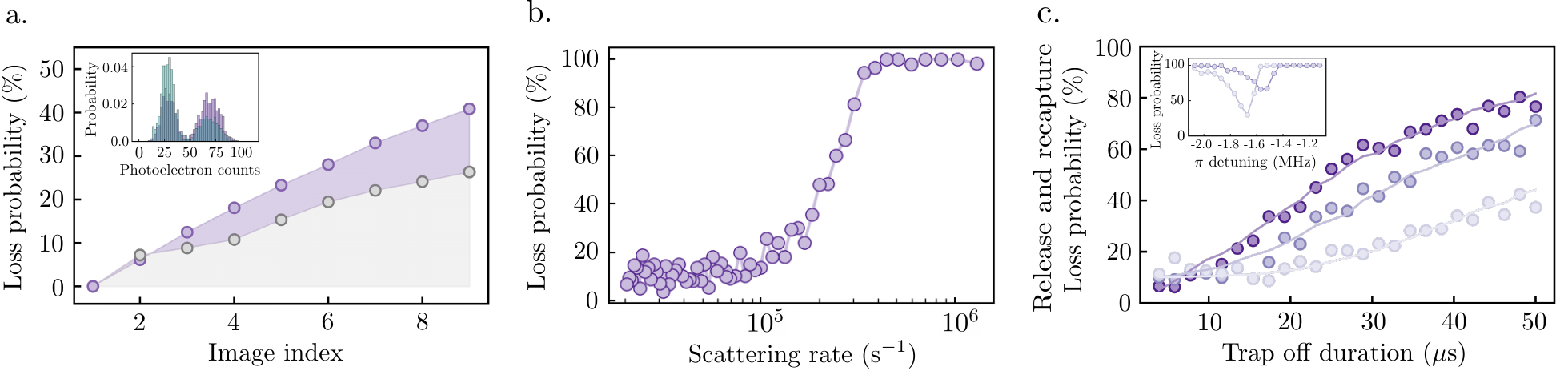}%
\caption{\textbf{Imaging-induced atom loss, heating, and cooling: a.} Probability of atom loss as a function of the number of consecutive $70$~ms-long images. The purple markers show total atom loss, while the gray markers indicate loss due to vacuum lifetime (in the absence of cooling/imaging light). The inset displays a histogram for the first (purple) and ninth (green) images of an atom in the same trap. \textbf{b.} Atom loss probability as a function of imaging beam scattering rate. The sudden increase when the scattering rate reaches twice the radial trapping frequency is due to heating from multiple photon recoils before the atom has had enough time to turn around in the trap. \textbf{c.} Release and recapture measurements obtained for EIT imaging (purple, $T = 16^{+4}_{-3}\,\mu$K) and pure EIT cooling without the imaging beam (gray, $T = 9^{+3}_{-2}\,\mu$K). For comparison we show the higher temperature from PGC in zero magnetic field (dark purple, $T = 37^{+30}_{-15}\,\mu$K). The inset displays atom loss as a function of EIT probe detuning, with and without the imaging beam.
}
\label{fig:Imaging Details}
\end{figure*}

For imaging to be effective in a continuously-operating array at a finite magnetic field, it is essential to maximize the atom survival probability. In Fig.~\ref{fig:Imaging Details}a, we measure the rate of atom loss by taking nine successive images, each for $70$~ms, corresponding to 
$\approx 6\times 10^3$ scattered photons per atom per image. The purple markers show the atom loss probability after $N$ images, while gray markers indicate the baseline atom loss, for which atoms are held in tweezers for the same duration without applying any cooling or imaging light. The graph indicates that the loss during imaging is mainly limited by the vacuum lifetime; however, after correcting for this loss, we find that the atom loss per image is $1.8(3)\,\%$, i.e. $0.026(4)\,\%/$ms or $3.0(5)\times 10^{-6}$ per scattered photon (for a discussion of a possible loss mechanism, see below). Our results imply that with an upgraded objective (an objective with NA = 0.65, $90\%$ transmission, and therefore imaging duration  $\approx10$~ms~\cite{zhang_high_optical_regal, sepehr_thesis, chiu2025continuousoperationcoherent3000qubit}), in combination with better vacuum lifetime, the atom loss during imaging could be reduced to the $10^{-3}$ level.

To investigate the speed limitation of our imaging technique, we scan the detuning of the fluorescence beam, thereby changing the scattering rate for fixed image duration. As shown in Fig.~\ref{fig:Imaging Details}b, we observe a sharp increase in atom loss for scattering rates above $\Gamma_{sc}\approx 10^5$ s$^{-1}$, 
which is on the order of twice the radial trapping frequency. This supports the hypothesis that the main limitation to the scattering rate is heating caused by multiple momentum kicks from the fluorescence beam in one direction before the atom has had enough time to turn around in the trap. This picture is confirmed by additional experiments (see Appendix~\ref{app:speedup}),
which find that increasing the trap depth enables higher scattering rates in proportion to the trapping frequency increase. Tighter tweezers with higher trapping frequencies generated via higher--NA objectives would allow for higher scattering rates without losing atoms, thereby enabling a further speedup of imaging. Furthermore, retroreflecting the imaging beam -- possibly frequency-shifted to avoid standing-wave effects -- could also enable higher scattering rates and shorter imaging times.

To further characterize the system, we measure the kinetic temperature of atoms in the radial direction with a standard release and recapture technique~\cite{tuchendler_pra} after EIT cooling with and without the additional heat load from the imaging beam, and compare it to PGC cooling in zero magnetic field. As shown in Fig.~\ref{fig:Imaging Details}c, a temperature of $T = 16^{+4}_{-3}\,\mu$K is achieved after applying one EIT imaging cycle, which is significantly lower than that achieved via PGC ($T = 37^{+30}_{-15}\,\mu$K), though it is higher than the EIT cooling temperature without the imaging beam ($T = 9^{+3}_{-2}\,\mu$K).
We note that to efficiently cool the atoms in the presence of the imaging beam, the probe beam frequency must be shifted by $-130$\,kHz compared to cooling in the absence of imaging (inset in Fig.~\ref{fig:Imaging Details}c), which is consistent with the differential AC Stark shift between magnetic sublevels due to the imaging beam (see Appendix~\ref{app:setup}).

\section{IV. LOSS MODEL}

In neutral-atom quantum computing experiments with quantum error correction, mid-circuit state detection~\cite{Norcia2023midcircuit, Lis2023midcircuit, dolev_logicalcomputer} must be performed many times, so it is important to understand and minimize the imaging loss.
The simplest model for atom loss during cooling is random diffusion to the exponential tail of the Boltzmann energy distribution, giving atoms a sufficiently large energy to leave the trap. However, this is inconsistent with the measured ratio of trap depth $U_0$ to temperature $T$ of $\eta = U_0/(k_B T) \gtrsim 50$~\cite{tuchendler_pra}, which gives exponentially small loss probability. A similar discrepancy persists when estimating the survival probability of imaging in other atom arrays (e.g.~\cite{covey_yb_imaging,manetsch2025,Norcia2023midcircuit,covey_2000_times_endres,zhang_high_optical_regal}), all with different experimental parameters and/or atomic species (see Table~\ref{tab:physics}).

\begin{figure}[ht!]
\centering
\includegraphics[width=0.5\textwidth]{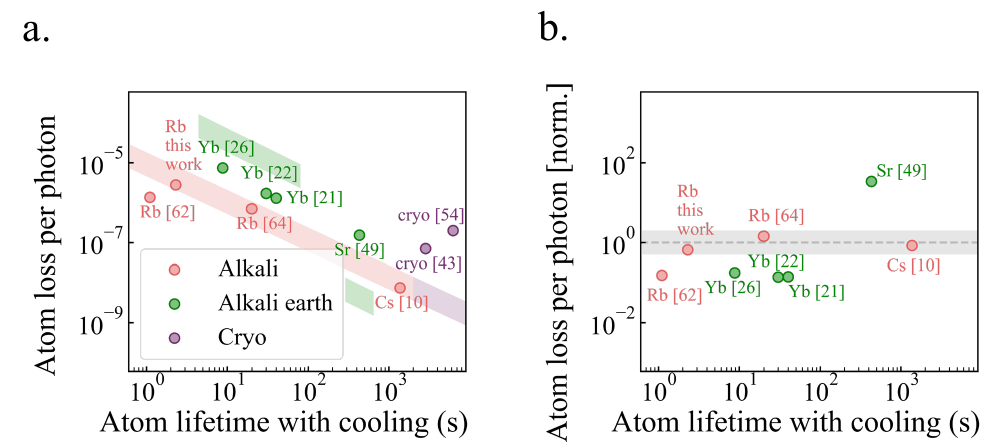}
\caption{\textbf{Imaging-induced atom loss across a variety of atom array experiments and atomic species.} Circles denote experimental observations, and shaded regions indicate predictions from our semi-quantitative loss model (see text), which shows order-of-magnitude agreement and validates our hypothesis of light-induced collisions. \textbf{a.} Atom loss probability per scattered photon as a function of atom lifetime, which shows that loss per scattered photon increases with shorter atom lifetime.
\textbf{b.} Atom loss per photon normalized by vacuum lifetime and properties of the atomic species according to our model (see Appendix~\ref{app:loss_model}),
which should collapse all experimentally-measured losses to one. Our model does not capture the behavior of alkaline-earth nor cryogenic experiments, where the background pressure is expected to be dominated by hydrogen.
Nevertheless, the model shows good agreement with the excess imaging loss observed across several alkali atom array experiments.
}
\label{fig:atom_loss}
\end{figure}

To explain the observed lifetime reduction during imaging, we postulate that the dominant loss mechanism is due to collisions between optically excited cold atoms and hot background gas atoms. These collisions have a significantly larger cross section than ground-state collisions due to off-resonant van der Waals interactions, resulting in an enhancement that scales with $(c/\overline{v})^{2/5}$:
\begin{equation}
    \frac{\overline{\sigma_{eg}}(T)}{\sigma_{gg}} \approx 1.57 \left( \frac{c}{\overline{v}} \right)^{2/5}
    \label{Eq:cross_section_ratio}
\end{equation}

where $\overline{v}$ is the RMS velocity of same-species background atoms. We assume that the background gas is dominated by same-species atoms, with a density ratio near unity across different experimental setups — often a reasonable assumption given the various loading methods (see Table~\ref{tab:physics}).
A second key factor is the fraction of time atoms spend in the excited state, set by the ratio of the photon scattering rate to the transition linewidth (see Appendix~\ref{app:loss_model} for details).

To validate our model, we analyze atom loss data from a range of experiments. A direct comparison of loss per scattered photon (Fig.~\ref{fig:atom_loss}a) is insufficient; once we account for differences in vacuum conditions and atomic species, we find reasonable agreement across a wide range of experiments involving room-temperature alkali atoms (Fig.~\ref{fig:atom_loss}b). 
We further validate our hypothesis with a separate Cs tweezer apparatus (see Appendix~\ref{app:loss_model}). We observe that when the Cs partial pressure is increased by a factor of 2, the bright lifetime decreases by a factor of 3, while the dark lifetime remains unchanged. These observations indicate that the background pressure of same-species atoms enhances atom loss from tweezers in the presence of near-resonant light, consistent with our hypothesis.
However, our model does not fully capture results from cryogenic setups~\cite{schymik_paper,zhang_high_optical_regal}, nor from alkaline-earth experiments~\cite{covey_2000_times_endres, covey_yb_imaging, Norcia2023midcircuit, Lis2023midcircuit} in which the same-species vapor pressure is expected to be low and additional loss mechanisms have been identified.

\section{V. CONCLUSION AND OUTLOOK}
In summary, we presented a simple and robust scheme to enable single-atom enhanced loading and non-destructive, high-fidelity imaging of atom arrays in a finite magnetic field. We show that our loading and imaging technique is capable of {more efficient} three-dimensional cooling to the temperatures necessary for high-fidelity qubit operations, rendering our method compatible with zoned architectures for neutral-atom quantum processors~\cite{dolev_logicalcomputer,bluvstein2025architecturalmechanismsuniversalfaulttolerant} and schemes for continuous reloading~\cite{RbCs_conveyorbelt,norcia2024iterative, zeiher_continuous,chiu2025continuousoperationcoherent3000qubit,li2025fastcontinuouscoherentatom}. Although the measurements presented here were performed at a bias field of $2.3\,\textrm{G}$ due to technical limitations, we have experimentally confirmed that our scheme can operate in a bias field of $10\,\textrm{G}$, which is similar to the one used for high-fidelity quantum gates in $^{87}$Rb~\cite{evered_high_fidelity} (see Appendix~\ref{app:10g} for more details).

Our results can be significantly improved with modest technical upgrades, in particular, an objective with higher NA and suitable AR coating. Objectives typically used in atom-array experiments (NA = 0.65 and $90\%$ imaging transmission) would enable an overall reduction in exposure time by a factor of $>10$, given the improvement in collection efficiency and tighter tweezer confinement, allowing higher photon scattering rates (see Appendix~\ref{app:speedup} and Table~\ref{tab:speedup}). The use of a glass cell instead of a steel science chamber is expected to decrease our background photon scattering due to improved optical access. Overall, we anticipate at least an order of magnitude reduction in imaging duration, as well as a similar reduction in single-atom preparation time. Switching to EIT cooling on the $D_1$ line, which has been used to achieve $\gtrsim 90\%$ single-atom loading probability \cite{brown_gray_molasses}, could further improve the enhanced single-atom loading probability.

The direct conclusion from our imaging-induced atom loss model is that improved vacuum lifetime could also reduce imaging-related atom loss. Additionally, our model implies that broader transitions contribute less imaging loss than narrow transitions for the same scattering rate, owing to the shorter time that atoms spend in the excited state. Given that many atom array experiments have similar vacuum environments, imaging and trapping parameters, and atom loading schemes, understanding the origin of the excess loss during imaging can help scale atom arrays to larger sizes. \\

\textit{Note added:} During the preparation of the manuscript we became aware of a related work, observing enhanced loading probabilities of $80(6)\%$ on the $D_2$ line of $^{87}$Rb in a single tweezer, albeit at zero magnetic field~\cite{Pampel2025quantifying}.

\section{ACKNOWLEDGEMENTS}
We would like to thank Simon Evered, Dolev Bluvstein, Simon Hollerith, Tout Wang, and Zhenpu Zhang for insightful discussions, Ashrit Verma for technical assistance, and Hongyi Meng and Marcello Laurel for careful proofreading of the manuscript. E.H.Q. acknowledges the support of the Natural Sciences and Engineering Research Council of Canada (NSERC). S.T. acknowledges support from the Rothschild fellowship of the Yad Hanadiv foundation, the VATAT-Quantum fellowship of the Israel Council for Higher Education, the Helen Diller Quantum Center postdoctoral fellowship and the Viterbi fellowship of the Technion – Israel Institute of Technology. This material is based upon work supported by the U.S. Department of Energy, Office of Science, National Quantum Information Science Research Centers, Quantum Systems Accelerator. Additional support is acknowledged from the NSF-funded Center for Ultracold Atoms (grant number PHY-2317134), the DARPA ONISQ program (grant number 134371-5113608), and the IARPA ELQ program (grant number W911NF2320219).

\appendix
\renewcommand{\thesection}{\Alph{section}}  %

\refstepcounter{appendixctr}\label{app:imaging_review}
\section{APPENDIX \theappendixctr: IMAGING PARAMETERS ACROSS TWEEZER ARRAY EXPERIMENTS}

In Table~\ref{tab:cooling}, we present a summary of cooling and non-destructive imaging methods for various atomic species and the associated parameters, including magnetic fields, trap depths, and photon scattering rates.

\begin{table*}[htbp!]
\centering
\footnotesize
\begin{tabular}{lccccccc}
\toprule
Atom & Cooling technique & Cooling dimensions & Magnetic field \hspace{1em} &  Trap depth & Temperature & Estimated $\Gamma_\textrm{sc}$ & Reference\\
 &  &  & G\hspace{1em} & [mK] & [uK] & [$10^5$ s$^{-1}$] & \\
\midrule

Rb & EIT & 3D & 2.3 & 0.88 & 16  & 0.9 & This work\\ 
Rb& Raman sideband & 3D & 3~\cite{kaufman_raman_cooling} & 1.1 & $\leq 8\ssymbol{1}$  & $\leq 0.07$ & \cite{cindy_coolingimaging}\\
Rb& Finite-field PGC & 1D (radial, $\leftrightarrow$) & 4.2 & 0.37 & 12  & 1 & \cite{chiu2025continuousoperationcoherent3000qubit}\\
Rb& Finite-field PGC  & 1D (radial, $\leftrightarrow$) & $< 8.6$ & 0.8 & 15  & 0.6 & \cite{dolev_transport,dolev_logicalcomputer,bluvstein2025architecturalmechanismsuniversalfaulttolerant} \\

Rb & Doppler & 3D ($\leftrightarrow$) & 20 & $\mathbf{8-10}$ & \textbf{100}  & 16 & \cite{saffman2017_state_selective}\\

Yb & Doppler & 1D (radial) & 500 & 0.35 & 4 & 1.5 & \cite{Norcia2023midcircuit}\\
Yb& Doppler & 2D (axial + radial) & 16 & 0.4 & 3  & 4 & \cite{Lis2023midcircuit} \\
Yb& Doppler & 2D (axial + radial, $\leftrightarrow$) & 58 & 0.58 & 5 & 1 & \cite{covey_yb_imaging}\\

\midrule 

Rb &  PGC & 3D ($\leftrightarrow$) & zero & 0.84 & 13  & 3.8 & \cite{zhang_high_optical_regal}\\
Rb& PGC & 3D ($\leftrightarrow$)  & zero & 1 & 50  & 2 & \cite{schymik_thesis}\\

Cs & PGC & 2D (radial, $\leftrightarrow$) & zero & 0.18 & 4.3  & 0.7 & \cite{manetsch2025} \\

Rb & $\Lambda$ Gray Molasses & 3D  & zero & 0.63 & 50  & - & \cite{brown_gray_molasses} \\

K & $\Lambda$ Gray Molasses  & 3D & zero & 1.1 & 20 & 1.4& \cite{gadway2022}\\

Li & $\Lambda$ Gray Molasses  & 3D & zero & 3.4 & 100 & 2.2 & \cite{Li_GrayMolassesImaging}\\

Sr & (Attractive) Sisyphus & 1D (radial) & zero & 0.45 & $< 5$ & 0.1 & \cite{covey_2000_times_endres}\\
Sr &  (Attractive) Sisyphus  &  1D (radial) & zero & 0.13 & $3$ & 0.23 & \cite{schreck_sr_tweezer}\\

\bottomrule
\end{tabular}
\caption{\textbf{Summary of non-destructive imaging parameters achieved across different tweezer array experiments.} For alkaline-earth and alkaline-earth-like atoms, the presence of narrow-linewidth transitions enables Doppler cooling in a finite magnetic field to reach temperatures of a few microKelvin. For alkali atoms, sub-Doppler cooling techniques (such as PGC or gray-molasses) are necessary, which require zero fields. Doppler cooling and imaging have been demonstrated in a finite field, however, significantly deeper traps were required to improve atom retention (in bold). 1D PGC in a finite field has been recently demonstrated, in which cooling is possible only along the magnetic field axis. On the other hand, Raman (EIT) cooling enables operation at a finite magnetic field, in which the two-photon detuning is carefully chosen to compensate the Zeeman splitting between ground hyperfine states. In the table, the cooling plane (radial, axial) is specified relative to the orientation of the optical tweezer. The use of retro-reflected or counter-propagating beams is indicated by ($\leftrightarrow$). When available, trap depths and temperatures are reported during imaging. The top half of the table indicates experiments operating in a finite magnetic field, the bottom half in zero magnetic field. $\ssymbol{1}$This temperature is inferred. 
}
\label{tab:cooling}
\end{table*} 

\refstepcounter{appendixctr}\label{app:setup}
\section{APPENDIX \theappendixctr: EXPERIMENTAL SETUP}

An array of optical tweezers is generated with a spatial light modulator (SLM; Hamamatsu x13138-02) focused through a microscope objective lens (Mitutoyo M Plan Apo NIR B 20X 378--867--5, $\textrm{NA}=0.4$). The experimental sequence consists of a 100-ms-long magneto-optical-trap (MOT) cooling and loading stage, followed by a 40-ms-long MOT compression stage~\cite{our_bec_paper}, during which an ensemble of $^{87}$Rb atoms is loaded into each tweezer. The magnetic gradient is then switched off, and the bias field along the $x$ direction is increased to $2.3$~G. All measurements presented in this work are performed in this constant magnetic field.

The $\sigma^+$--polarized EIT control beam addresses the $\ket{5S_{1/2}, F=2, m_F=1} \rightarrow \ket{5P_{3/2}, F'=2, m_F'=2}$ transition with measured Rabi frequency of $\Omega_c/(2\pi) = 7.5\,$MHz. For the single-photon detuning of $\Delta_c/(2\pi) = +36\,$MHz used for cooling (relative to the trap-shifted line), the estimated AC Stark shift of the control light is given by $\delta_\textrm{AC}/(2\pi) = \Omega_c^2/4\Delta_c = 390\,$kHz. The expected Fano linewidth $\gamma_{2\textrm{ph}}/(2\pi) =  2\delta_\textrm{AC} (\Gamma/\Delta_c) = 130\,$kHz is similar to the observed linewidth of $\approx 150\,$kHz, which is
larger than the trapping frequencies ($\omega_r/(2\pi)= 45$\,kHz and $\omega_z/(2\pi)= 4$\,kHz). We note that the observed Fano linewidth does not decrease with lower laser power, and that we cannot resolve sidebands corresponding to motional  transitions. For our tweezers, the Lamb-Dicke parameter is $\{\eta_r, \eta_z\} \approx \{0.29, 0.97\}$ and our lowest attainable temperature of $\approx 9\,\mu$K with this cooling method (average occupation numbers $\{\bar{n}_r, \bar{n}_z\} \approx \{3.7, 46.4\}$) indicates we work marginally in the Lamb-Dicke regime in the radial direction, and well outside the Lamb-Dicke regime in the axial direction.

The $\pi$--polarized EIT probe beam addresses the $\ket{5S_{1/2}, F=2, m_F=2} \rightarrow \ket{5P_{3/2}, F'=2, m_F'=2}$ transition. The probe beam aligned in the $y$--$z$ plane has measured Rabi frequency of $\Omega_p/(2\pi) = 0.52~$MHz, while the probe light aligned along the $y$ direction has $\Omega_p/(2\pi) = 0.79$\,MHz. The ratio between the two beams was chosen by maximizing the imaging signal-to-noise ratio.

The $\sigma^+$--polarized fluorescence imaging beam is $-67\,$MHz detuned (relative to the trap-shifted line) from the $\ket{5S_{1/2}, F=2, m_F=2} \rightarrow \ket{5P_{3/2}, F'=3, m_F'=3}$ transition. We estimate the scattering rate of this beam to be $\Gamma_{sc} \sim 10^5\,$ s$^{-1}$ from the laser power and beam waist, roughly consistent with the $-130(10)\,$kHz AC Stark shift shown in the inset of Fig.~\ref{fig:Imaging Details}c of the main text. A polarized repump beam addresses the $\ket{5S_{1/2}, F=1} \rightarrow \ket{5P_{3/2}, F'=2}$ transition, and is generated as a sideband on the $\sigma^+$ control beam via a phase electro-optic modulator (EOM; iXblue NIR-MPX800-LN-05). The parameters of our cooling and imaging scheme are summarized in Table~\ref{tab:parameters}.

\begin{table}[H]
\centering
\footnotesize
\begin{tabular}{lccc}
\toprule
Laser beam & Transition & $\Omega/2\pi$ [\textrm{MHz}] & $\Delta/2\pi$ [\textrm{MHz}] \\
\midrule
EIT $\sigma^+$ control & $\ket{F=2} \rightarrow \ket{F'=2}$ & 7.5 & 36  \\
EIT $\pi$ oblique probe & & 0.52 & $36$ \\
EIT $\pi$ radial probe & & 0.79 & $36$ \\
Fluorescence $\sigma^+$ & $\ket{F=2} \rightarrow \ket{F'=3}$ & 5.9 & $-67$ \\
\bottomrule
\end{tabular}
\caption{\textbf{Summary of parameters used for EIT imaging.} All laser beams address transitions on the $^{87}$Rb $D_2$ line, where $F$ and $F'$ denote the hyperfine ground and excited states, respectively. The detunings are reported relative to the trap-shifted resonances.}
\label{tab:parameters}
\end{table}

\refstepcounter{appendixctr}\label{app:imaging_details}
\section{APPENDIX \theappendixctr: EIT IMAGING DETAILS}

\begin{figure*}[t!]
\centering
\includegraphics[width=\textwidth, trim= 0 0 0 0,clip]{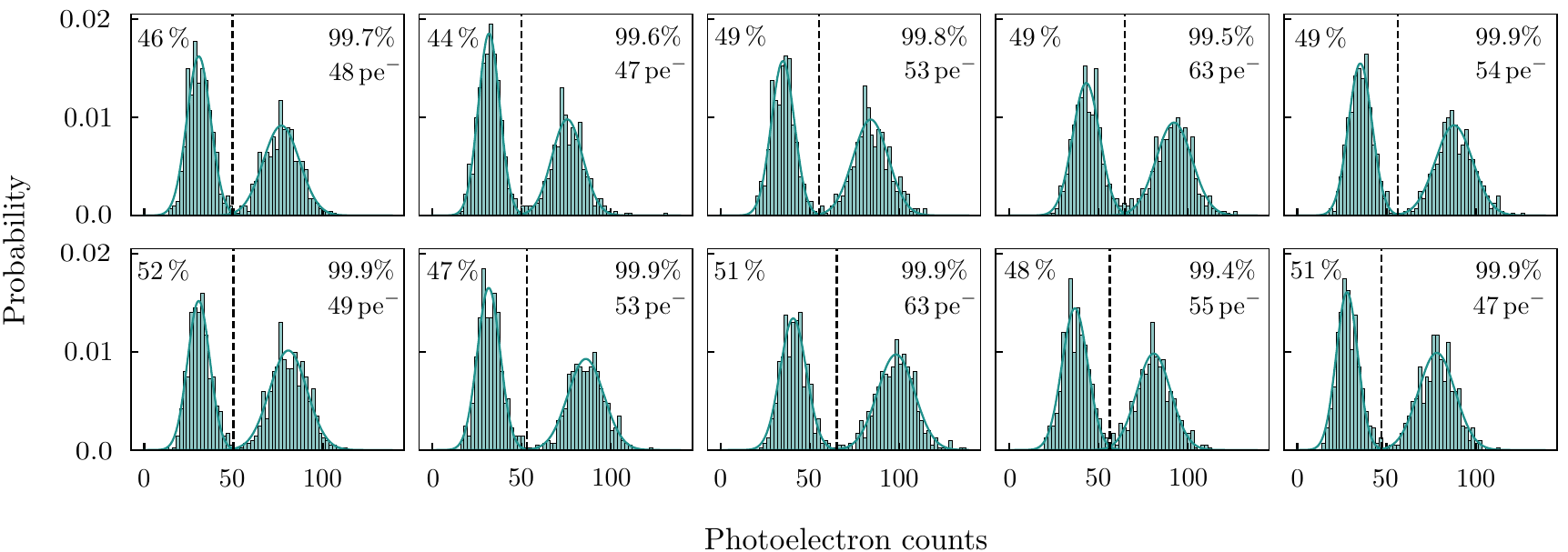}
\caption{\textbf{EIT imaging in a $2.3$\,G magnetic field:} Details pertaining to the imaging results shown in the main text (Fig.~\ref{fig:Imaging and Loading}b) are presented here for each trap, including the imaging fidelity and threshold (in photoelectron counts, pe$^-$) shown in the upper right of each histogram, and the single-atom loading probability in the upper left.}
\label{fig:2g_histograms}
\end{figure*}

Our imaging results presented in the main text (Fig.~\ref{fig:Imaging and Loading}b) are shown in detail in Fig.~\ref{fig:2g_histograms}, with histograms, fidelities, and thresholds listed per trap. The camera counts are summed over a $4\times4$ pixel area around each atom, and a constant dark offset is subtracted. This offset (200 counts on average) pertains to the properties of our quantitative CMOS camera (Hamamatsu C15550-20UP ORCA-Quest). The net counts are then multiplied by the conversion factor (0.11 electrons/count) to estimate photoelectron counts ($x$--axis of all histograms presented in this manuscript), and then divided by the quantum efficiency ($\approx 0.5$ at 780\,nm) to estimate the number of photons incident on the camera. Each histogram is fitted to the sum of two Gaussian distributions (corresponding to zero and one atom), and the overlap integral between the fits gives the imaging infidelity.

\refstepcounter{appendixctr}\label{app:speedup}
\section{APPENDIX \theappendixctr: IMAGING SPEEDUP WITH MODEST TECHNICAL UPGRADES}

We expect that our imaging duration of $70~$ms can be significantly reduced with modest technical upgrades. Such a speedup is crucial for fast mid-circuit readout in state-of-the-art, continuously-operating neutral atom quantum processors. The imaging duration is expected to scale as:
\begin{equation}
    T_\textrm{img} = 70\,\textrm{ms}\, \cdot \frac{60\%}{\textrm{T}_{780}} \cdot \frac{3.3 \%}{\eta}\cdot \frac{2\pi \, \cdot 4\,\textrm{kHz}}{\omega_z}
\end{equation}

where $\textrm{T}_{780}$ is the objective transmission at 780\,nm, $\eta = \frac{3}{16}\left(\frac{8}{3}-3x + \frac{x^3}{3}\right)$ and $x = \sqrt{1-\textrm{NA}^2}$, where $\eta$ is the collection efficiency, defined as the fraction of scattered photons collected by the objective for a dipole driven by circularly-polarized light when the quantization axis is orthogonal to the objective axis. In Table~\ref{tab:speedup}, we outline the different contributions, leading to a speedup of $4.5$ by simply switching to a high-NA, suitably-coated microscope objective which collects ($\eta$) and transmits ($\textrm{T}_{780}$) a larger fraction of the scattered photons. Additionally, tighter confinement -- especially along the axial direction -- of atoms in tweezers for the same trap depth should enable an increase in scattering rate proportional to the axial trap frequency $\omega_z$, which scales as $w_0^{-2}$ for fixed trap depth where $w_0$ is the tweezer waist. Overall, we expect a speedup of over an order of magnitude, which would reduce the necessary exposure time to $\lesssim 10\,$ms for the same imaging fidelity.

To confirm that the scattering rate during imaging is limited by the trapping frequency, we increase the trap depth by $\approx 40\%$, corresponding to an increase in trapping frequency of $\approx 18 \,\%$. For the same image duration, we find that the scattering rates at which we retain $50\%$ survival probability increases by $\approx 23 \,\%$, in good agreement with the increase in trapping frequency, rather than the trap depth. This implies that for the same trap depth, the scattering rate can be increased with better tweezer confinement, as expected.

\begin{table}[htbp]
\centering
\footnotesize
\begin{tabular}{lccc}
\toprule
Parameter & Current & Projected & Speedup \\
& value & value & factor\\
\midrule
Numerical aperture & 0.4 & 0.65 & -- \\
Collection efficiency $\eta$ & 3.3$\%$ & 10.0$\%$& 3\\
Transmission $\textrm{T}_{780}$ & 60$\%$ & 90$\%$ & 1.5\\
\midrule
Tweezer waist $w_0$ & 1.6\,$\mu$m & $1\, \mu$m & -- \\
Axial trap frequency $\omega_z/2\pi$ \hspace{5pt} & 4 kHz & 10 kHz & 2.5 \\
\midrule
\midrule
& & \textbf{Speedup} & $\approx$ 11.3 \\
\bottomrule
\end{tabular}
\caption{\textbf{Summary of parameters and speedup factors expected from an upgraded microscope objective.} The collection efficiency is the fraction of scattered photons captured by the objective, for a dipole driven by circularly-polarized light with the quantization axis orthogonal to the objective axis. The improvement in axial trap frequency is estimated based on its dependence on tweezer waist for Gaussian beams ($\sim w_0^{-2}$) for fixed trap depth.}
\label{tab:speedup}
\end{table}

\begin{figure*}[ht!]
\centering
\includegraphics[width=\textwidth, trim= 0 0 0 0,clip]{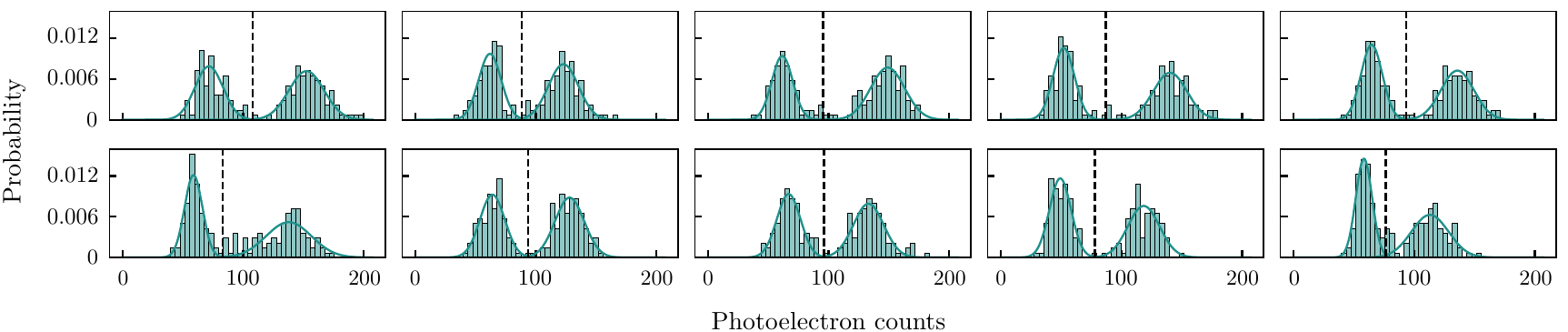}
\caption{\textbf{EIT imaging in a $10$\,G magnetic field:} Histograms are taken for $150$\,ms of exposure time, with imaging fidelities ranging from $98.4\%-99.9\%$. A detailed study at these fields (including fidelities, atom losses, and loading rates) remains the subject of further investigation.}
\label{fig:10g_histograms}
\end{figure*}

State-of-the-art atom array experiments observe backgrounds of $ < 0.4$ photons per scattered photon by an atom~\cite{bluvstein2025architecturalmechanismsuniversalfaulttolerant, chiu2025continuousoperationcoherent3000qubit, manetsch2025}, however, we find that our oblique EIT $\pi$ cooling beam contributes significantly to our background counts ($\approx 0.7$ ph/ph) due to the geometry and limited optical access of our vacuum chamber. This beam provides axial cooling, and its projection along the objective axis coupled with its beam path lead to large backgrounds. By cooling instead on the $D_1$ line while imaging on the $D_2$ line, scattered $D_1$ photons can be spectrally filtered at the camera, thereby further reducing both the background photon number and the imaging duration. Alternatively, the background scattering could be reduced by switching to a glass-cell-based apparatus with improved optical access.

In the future, we envision that our method will be employed in continuously-operated atom arrays, where angled beams with momentum components along all three axes could be used to perform 3D cooling of the atoms. Both beams can be focused to accommodate the zone architecture, as demonstrated in~\cite{bluvstein2025architecturalmechanismsuniversalfaulttolerant}, and state-selective imaging can be performed via a state-selective lattice enabling loss detection~\cite{bluvstein2025architecturalmechanismsuniversalfaulttolerant}. Since our method can cool both the transverse and the axial degrees of freedom, it could yield lower temperatures in deep quantum circuits with several rounds of mid-circuit measurement, as well as improved survival rates for shorter imaging durations. That said, further study is necessary to ascertain the performance of our technique in tighter tweezers and lower background levels.

\refstepcounter{appendixctr}\label{app:10g}
\section{APPENDIX \theappendixctr: EIT IMAGING IN A 10\,G MAGNETIC FIELD}

Due to a technical limitation at the time of this work, it was not possible to stabilize magnetic bias coil currents larger than $1$\,A along the radial axis of our tweezers, which limited the bias field to $\leq2.3$\,G. After upgrading our current controller, $10$\,G fields were possible and a brief measurement was carried out. As shown in Fig.~\ref{fig:10g_histograms}, EIT imaging of single atoms on the order of 99\% fidelity is possible at the magnetic fields necessary for high fidelity quantum gates. We note, however, that a detailed study of imaging fidelities, atom losses, and loading rates at such fields is the subject of further investigation.

\refstepcounter{appendixctr}\label{app:application}
\section{APPENDIX \theappendixctr: APPLICATION TO CONTINUOUSLY-RELOADED ATOM ARRAYS}

Several recent demonstrations of continuously-operated atom arrays~\cite{RbCs_conveyorbelt,norcia2024iterative,chiu2025continuousoperationcoherent3000qubit,li2025fastcontinuouscoherentatom} feature two-chamber systems, in which a dipole trap or optical conveyor belt delivers pre-cooled atomic ensembles from a MOT chamber to a science chamber, where they serve as an atomic reservoir under a static magnetic field. Moving tweezers then repeatedly extract atoms from the reservoir to replenish empty sites in an array of single atoms. In the present work, which operates in a single chamber, atoms are first loaded into tweezers from a MOT, after which the magnetic field gradient is switched off and a static bias field is applied. At this stage, each tweezer is occupied by a small ensemble of atoms (evident in Fig.~\ref{fig:Imaging and Loading}a of the main text), and our technique demonstrates that laser cooling, parity-projection, and non-destructive imaging can all be performed without the need for subsequent field switching. Therefore, our technique is directly applicable to continuously-reloaded atom arrays, enabling non-destructive imaging of atoms in readout zones, as well as parity-projection of ensembles extracted from the reservoir, under a truly static field environment.

\refstepcounter{appendixctr}\label{app:loss_model}
\section{APPENDIX \theappendixctr: IMAGING-INDUCED ATOM LOSS MODEL}

In this section, we provide a simple semi-quantitative model that can explain the enhanced atom loss during imaging observed in many atom array experiments, as well as our own. We observe that in many experiments, the loss rate during imaging scales with the background loss rate when the atoms are trapped in the dark (see Table~\ref{tab:physics}). The assumption of this model is that the excess loss stems from enhanced interactions between electronically-excited cold atoms and hot (background) atoms of the same species. In this case, due to near-resonance dipole-dipole interactions, the collision cross section between an optically excited and a ground-state atom is much larger than for collisions between two atoms in their ground states, or between a trapped atom and some other background gas atom or molecule~\cite{bjorkholm_collision_limited_1988}.

Ground-state atoms in tweezers interact with background gas atoms via van der Waals interactions $-C_{6, gg}/r^6$, where $r$ is the interatomic distance and the interaction coefficient is: 
\begin{equation}
C_6 = \frac{3e^4\hbar}{2m_e^2} \sum_{i_a,i_b} \frac{f_{ig}^a f_{ig}^b}{\omega_{ig}^a \omega_{ig}^b (\omega_{ig}^a + \omega_{ig}^b)}.
\end{equation}
Here $e, m_e$ are the electric charge and mass of the electron, $i_a, i_b$ denote the excited states of atoms $a$ and $b$, and $f_{ig}^a, f_{ig}^b$ and $\omega_{ig}^a, \omega_{ig}^b$ are the oscillator strengths and frequencies of the transitions, respectively. If we assume that there is a dominant transition from the ground state to some excited state (e.g. for $^{87}$Rb, $5S_{1/2} \rightarrow 5P_{3/2}$) and take the oscillator strength to be one, the expression can be simplified to $C_{6, gg} = \frac{3}{4} \hbar \omega \alpha_0^2$~\cite{cohen_tannoudji_atom_photon_1998, mitroy_bromley_2003_vdw_coefficients}, where $\alpha_0$ is the static ground state polarizability and $\omega$ is the transition frequency.

Atoms in the excited state interact more strongly with atoms in the background gas of the same species, and for near-zero-velocity background gas atoms, the resonant interaction takes the form $C_{3, eg}/r^3$ and $C_{3, eg} \sim \hbar \Gamma/(2k^3)$, where $k=\omega/c$ and $\Gamma$ is the linewidth of the excited state. For a background gas atom at room temperature, Doppler broadening results in a larger energy defect, recovering again the form $C_{6, eg}/r^6$. Assuming a Maxwell-Boltzmann distribution of the background gas, only a negligible fraction of atoms have near-zero velocity. Therefore, we need only calculate how atoms interact in the non-resonant case. The Hamiltonian of one cold atom $a$ and one background atom $b$ can be written in the basis $\ket{e_{a}, g_{b}}, \ket{g_{a}, e_{b}}$:
\begin{equation}
H = 
\begin{bmatrix}
E^a_e + E_g^b & V_1 \\
V_1 & E^a_g + E_e^b
\end{bmatrix}
\end{equation}
where $E^a_{e/g},E_{e/g}^b$ are the energies of atoms $a$ and $b$ in the ground or excited state and $V_1 = C_{3,eg}/r^3$. In the regime of non-resonant interactions, the energy defect due to Doppler broadening is $\delta E \equiv |(E^a_e - E_g^a)-(E^b_e - E_g^b)| = k \cdot \overline{v} \gg |V_1|$, and from the eigenenergies we find:
\begin{align}
    C_{6, eg} & = \frac{2 \omega}{k \overline{v}} \, C_{6, gg}\\
& = \frac{2c}{\overline{v}} \, C_{6, gg}
\end{align}
where $\overline{v}$ is the background atom root-mean-square (RMS) velocity and $c$ is the speed of light. This implies that cold atoms in the excited state interact with hot atoms in the background gas with an enhancement factor $2c/\overline{v}$ compared to ground-ground interactions, which is on the order of $10^6$.

With this, the collision cross section due to an interaction of the form $C_{s}/r^s$ can be approximated as \cite{massey_free_paths_mohr,bernstein_semiclassical_1963,croucher_total_collision_clark}:
\begin{align}
     \sigma_{eg} &= \pi \, \frac{2s-3}{s-2} f(s)^{\frac{2}{s-1}} \cdot \left(\frac{4 C_s}{\hbar v}\right)^{\frac{2}{s-1}} \nonumber\\
     & = \left\{
     \begin{aligned}
         & 12\pi \cdot \frac{C_3}{\hbar v} \ && (s = 3) \\ 
         & 9.96 \cdot \left(\frac{C_6}{\hbar v}\right)^{\frac{2}{5}} && (s = 6) \\ 
     \end{aligned}
     \right.
     \nonumber
\end{align}

\begin{table*}[t]
\centering
\footnotesize
\begin{tabular}{lccccccccccccl}
\toprule
Atom & 
$U/k_B$ & 
$T\ssymbol{1}$  & 
$\Gamma_{\textrm{sc}}$& 
Loss$_{\text{img}}$& 
$T_{\text{img}}$ &
$\tau_{\text{img}} \ssymbol{2}$& 
$\tau_{\text{vac}}$ & 
\textbf{$\tau_{\text{img}}/\tau_{\text{vac}}$} & 
\textbf{$\tau_{\text{img}}/\tau_{\text{vac}}$} & 
Loading method & 
Reference \\
 & [mK] & [$\mu$K] & [$10^{5}\,$s$^{-1}$] & $\times 10^{-4}$ & [ms] & [s] & [s] & 
 Experiment & Theory &  & \\
 \midrule
$^{87}$Rb & 0.88 & 16 & 0.9 & 480 & 70 & 1.4 & 2.3 %
& \textbf{0.61} & \textbf{0.53} & background vapor+3D MOT & This work\\
$^{87}$Rb & 0.87 & - & 2 & 590 & 50 & 0.85 & 1.1 %
& \textbf{0.77} & \textbf{0.28} & background vapor+3D MOT & \cite{beili_cavity, private_communication}\\
$^{87}$Rb & 1 & 50 & 1$\ssymbol{4}$ & 60 & 50 & 8 & 20 %
& \textbf{0.40} & \textbf{0.51} & Zeeman slower+3D MOT & \cite{analysis_imperfections}\\
$^{133}$Cs & 0.18 & 4.3& $0.7$ & 1 & 80 & 763 & 1374 %
& \textbf{0.55} & \textbf{0.54} & 2D+3D MOT & \cite{manetsch2025}\\
\midrule
\blue{$^{87}$Rb (cryo)} & \blue{0.84} & \blue{13} & \blue{$3.8$} & \blue{3.8} & \blue{14} & \blue{37} & \blue{2850} %
& \blue{\textbf{0.01}}& \blue{\textbf{[0.16]}}& \blue{2D+3D MOT} & \cite{zhang_high_optical_regal, private_communication}\\
\blue{$^{87}$Rb (cryo)} & \blue{1} & \blue{50} & \blue{2}  & \blue{20} & \blue{50} & \blue{25} & \blue{6260} %
& \blue{\textbf{0.004}} & \blue{\textbf{[0.26]}} & \blue{Zeeman slower+2D+3D MOT} & \cite{schymik_thesis, private_communication}\\
\midrule
$^{88}$Sr & 0.45 & $<5$& $0.1$ & 6.8 & 50 & 73 & 431 %
& \textbf{0.17} & \textbf{0.93}& Zeeman slower+2D+3D MOT & \cite{covey_2000_times_endres}\\
$^{171}$Yb & 0.58& 5 & $1.0 $ & 100$\ssymbol{2}$ & 12 & 1.2 & 8.8$\ssymbol{3}$ %
& \textbf{0.14}& \textbf{0.03}& 2D+3D MOT& \cite{covey_yb_imaging}\\
$^{171}$Yb & 0.4 & 5 & 4 & 19 & 3.5 & 1.8  & 40 %
& \textbf{0.04} & \textbf{0.01} & 2D+3D MOT & \cite{Lis2023midcircuit, private_communication}\\
$^{171}$Yb & 0.35 & 4 & $1.5$& 75$\ssymbol{2}$ & 26 & 3.5  & 30 %
& \textbf{0.03}& \textbf{0.02} & optical lattice & \cite{Norcia2023midcircuit}\\
\bottomrule
\end{tabular}
\caption{\textbf{Excess imaging loss for various optical tweezer array setups:} $U$ is the trap depth during imaging, $T$ is the temperature, $\Gamma_\textrm{sc}$ is the estimated scattering rate, Loss$_{\text{img}}$ is the atom loss probability for duration $T_{\text{img}}$ under illumination of imaging light, $\tau_{\text{img}}$ is the lifetime ($1/e$) under continuous imaging (which is estimated by dividing the imaging duration $T_{\text{img}}$ by the imaging loss Loss$_{\text{img}}$, including vacuum loss), and $\tau_{\text{vac}}$ is the vacuum-limited tweezer lifetime under periodic (or continuous) cooling. The rows in blue are for cryogenic experiments where our theoretical model does not apply because the background vapor pressure of the trapped atom species is negligibly low. When available, the quoted temperature is after imaging, as an upper bound.
\\
$\ssymbol{2}$ For state-selective imaging, we take the (average) loss figure corresponding to the bright state(s).\\
$\ssymbol{3}$ The lifetime in the tweezer is reported without periodic (nor continuous) cooling. \\
$\ssymbol{4}$ The scattering rate is assumed to be $10^5$ s$^{-1}$ when unavailable. 
}
\label{tab:physics}
\end{table*}

We assume that a collision event with a room-temperature atom results in loss from the trap. We can then write the atom loss rate during imaging as:
\begin{align}
     \Gamma_{\text{loss,imaging}}&\approx \frac{\Gamma_{\text{sc}}}{\Gamma}  \,  n_\textrm{atom} \overline{\sigma_{eg}}(T)  \overline{v}\nonumber + \Gamma_{\text{loss,vac}}\\
     &= \frac{\Gamma_{\text{sc}}}{\Gamma}   \left( n_\textrm{tot} \sigma_{gg} \overline{v}\right ) \frac{n_\textrm{atom}}{n_\textrm{tot}} \frac{\overline{ \sigma_{eg}}(T)}{ \sigma_{gg}} \nonumber + \Gamma_{\text{loss,vac}}\\
     &=  \frac{\Gamma_{\text{sc}}}{\Gamma} \, \Gamma_{\text{loss,vac}} \frac{n_\textrm{atom}}{n_\textrm{tot}} 
         \frac{\overline{ \sigma_{eg}}(T)}{ \sigma_{gg}}  + \Gamma_{\text{loss,vac}}
         \label{Eq:ImagingLoss}
\end{align}
where $n_\textrm{atom}$ and $n_\textrm{tot}$ are the densities of the same-species background gas and the total background gas, respectively, $\overline{v}=\sqrt{3k_B T/m}$ is the RMS velocity of the background gas ($m$ and $T$ are the atomic mass and temperature of the background gas atom), $\overline{\sigma_{eg}}(T)$ is the average, temperature-dependent, excited-ground collision cross section, $\sigma_{gg}$ is the ground-ground collision cross section, $\Gamma_{\text{sc}}$ is the photon scattering rate, and $\Gamma_{\text{loss,vac}} = n_\textrm{tot} \sigma_{gg} \overline{v}$ is the atom loss rate in the dark, assuming that the ground-ground collision cross section is of the same order of magnitude for different species. The factor $\Gamma_{\text{sc}}/\Gamma$ accounts for the fraction of time spent by the cold atom in the excited state during imaging, during which it can participate in an excited-ground collision event with enhanced cross section.

As discussed previously, the excited-ground collision cross section is dominated by the contribution from the off-resonant $C_6$ interaction, given by:
\begin{align}
     \frac{\overline{\sigma_{eg}}(T)}{\sigma_{gg}} &= \frac{4}{\sqrt{\pi}} \int_0^{\infty}  dv \,  e^{-\frac{v^2}{v_p^2}} \frac{\sigma_{eg}(v)}{\sigma_{gg}} \frac{v^2}{v_p^3} \nonumber\\
     &\approx 1.57 \, \left(\frac{c}{\overline{v}} \right)^{2/5} 
         \nonumber
\end{align}

where $v_p=\sqrt{2k_B T/m}$ is the most probable speed of background gas atoms, and we assume that $n_\textrm{atom}/n_\textrm{tot}$ is of order unity. Therefore, the imaging loss rate $\Gamma_{\text{loss,imaging}}$ is enhanced relative to the vacuum loss rate $\Gamma_{\text{loss,vac}}$ by a factor $\alpha$:
\begin{align}
\alpha=\frac{\overline{\sigma_{eg}}(T)}{\sigma_{gg}}\frac{\Gamma_{sc}}{\Gamma}
\end{align}

For interactions between atoms and/or molecules of different species, the energy mismatch for both the excited and ground states of the cold atom is, \textit{a priori}, of similar magnitude. Consequently, the resulting $C_6$ coefficient remains comparable to that of ground-ground interactions and does not benefit from the $c/\overline{v}$ enhancement factor. Therefore, interactions between an excited-state cold atom and a ground-state background atom or molecule of a different species can be safely neglected as long as the partial pressure of the same species relative to the total background pressure remains substantially larger than $\left(\overline{v}/c\right)^{2/5} \approx 10^{-3}$.

The relevant parameters are summarized for all experimental setups in Table~\ref{tab:physics}. Surprisingly, our simple model explains the results of room-temperature alkali atom array experiments, regardless of the atom species. Since many of the above experiments utilize a 2D MOT, residual atomic flux from the 2D MOT may serve as a source for hot background atoms of the same species, supporting our assumption that collisions with same-species atoms dominate the loss mechanism. The same can also be said of residual background atoms from the 3D MOT. However, our model does not accurately describe alkaline-earth experiments, since the same-species vapor pressure is expected to be low. Additional loss mechanisms, such as trap-induced photoionization and Raman scattering out of the metastable manifold, have been identified~\cite{thompson_highfidelity, Lis2023midcircuit, Norcia2023midcircuit, covey_yb_imaging} and appear to explain the imaging loss. Similarly, our model should not be applied to the Rb cryogenic experiments~\cite{zhang_high_optical_regal,schymik_paper,schymik_thesis}, where the background vapor pressure of Rb should be low. Indeed, such systems observe significantly higher imaging losses than we predict.

We further validate our understanding by performing measurements in a different experimental apparatus consisting of $^{133}$Cs atoms trapped in tweezers within a bow-tie cavity~\cite{wang2025programmablefewatombraggscattering, matt_cavity}. The use of an adjustable angle valve on the atomic source enables control of the cesium partial pressure within the vacuum chamber. By probing the cavity linewidth, the background Cs partial pressure can be estimated due to the coupling between hot background atoms and cavity photons, which broadens the cavity transition. We find that when the cesium background pressure increases by a factor of $\approx$ 2 (indicated by a corresponding increase in the cavity linewidth), the bright tweezer lifetime decreases by a factor of 3, while the dark lifetime remains unchanged (the deviation from linear scaling is likely due to the limited accuracy of the measurement of partial pressure). This observation confirms that higher same-species background pressure enhances imaging-induced atom loss and supports our theoretical framework.

\bibliography{reference}

\end{document}